\documentclass[smallextended]{svjour3}       % onecolumn (second format)
\smartqed  % flush right qed marks, e.g. at end of proof
\usepackage{graphicx}
\begin{document}

\title{N$^*$ Experiments and their Impact on Strong QCD Physics}
\author{Volker D. Burkert \\ for the CLAS collaboration}
\institute{Volker D. Burkert \\
Thomas Jefferson National Accelerator Facility\\
Newport News, Virginia 23606, USA\\
              \email{burkert@jlab.org}   
}

\date{Received: date / Accepted: date}
\date{\today}

\maketitle

\begin{abstract}
I give a brief report on experimental studies of the spectrum and the structure of the excited states of the 
nucleon and what we learn about their internal structure. The focus is on the effort to obtain a more complete 
picture of the light-quark baryon excitation spectrum employing electromagnetic beams, and on the 
study of the transition form factors and helicity amplitudes an their dependence on the size of the 
four-momentum transfer $Q^2$, especially on some of the most prominent resonances.   
These were obtained in pion and eta electroproduction experiments off proton targets. 

\end{abstract}

\section{Introduction}
\label{intro}
For this introductory talk the organizers asked to address what we learn about strong QCD (sQCD) from the
 study of nucleon resonances transitions. Nathan Isgur said in the concluding talk at 
 N*2000:  "{\it I am convinced that completing this chapter in the history of science will be one of the 
 most interesting and fruitful areas of physics for at least the next thirty 
 years.}"       

We begin this conference in excited anticipation of tomorrow's solar eclipse, which, thanks to the organizers 
schedule, coincides with the second day of this conference. It allows me to refer to another famous Solar Eclipse 
of March 29, 1919, when Sir Arthur Eddington performed the first experimental test~\cite{Dyson:1920cwa} 
of Albert Einstein's general theory of relativity~\cite{Einstein:1916vd}. The findings led to the eventual triumph 
of general relativity over classical Newtonian physics. It also gave birth to modern scientific cosmology and the 
study of the history of the universe. 

\begin{figure}[t]
\centering\includegraphics[width=11.0cm,height=6.0cm,clip]{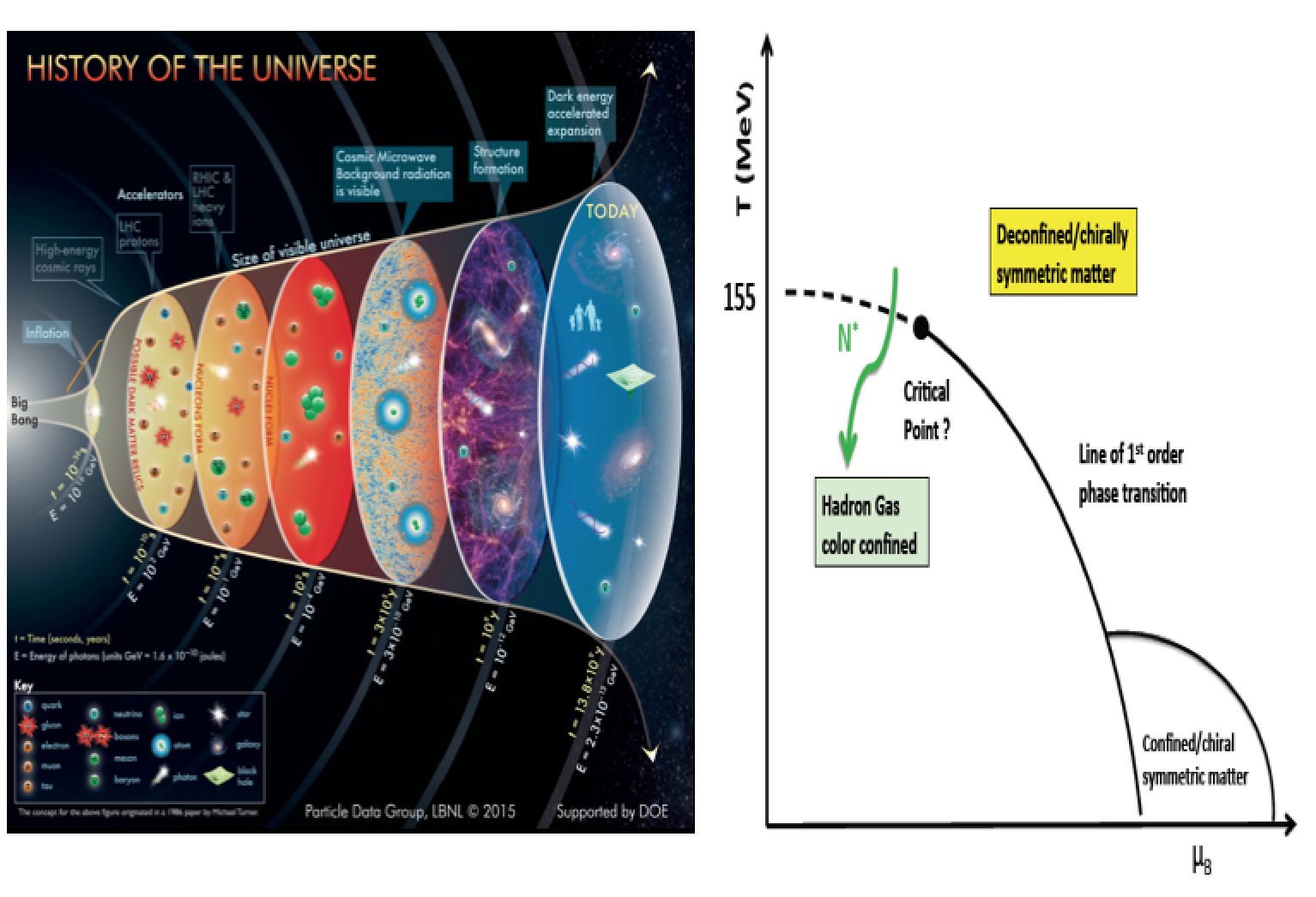}
\caption{Left panel: The evolution of the Universe as depicted by the LBNL Particle Data Group, 2015. 
The area characterized by the second disk from the left is where hadrons of confined quarks and gluons 
occur. The CEBAF electron accelerator has the energy reach to access this region and study 
processes in isolation that occurred in the microsecond old universe and resulted in the freeze 
out of baryons. Right panel: A generic phase diagram for the transition from the de-confined quark-gluon 
state to the confined hadron state.}
\label{universe}      
\end{figure}

In this meeting we also address how excited states of the nucleon fit into our understanding of the 
forces and the dynamics of matter in the history of the universe and in its current state. The Particle Data 
Group issues the beautiful representations of the phases through which the universe evolved from the 
Big Bang (BB) to our times as shown in Fig.~\ref{universe}.  
There are some marked events that have been of particular significance during the early phases of  
its history, such as the quark-gluon plasma (QGP) of non-interacting color quarks and gluons, the 
forming of nucleons, and of light nuclei. What is not shown, but is of particular significance for our field, 
is the transition from the QGP to stable nucleons that begins just microseconds after the BB, when 
dramatic events occurred - chiral symmetry is broken, quarks acquire mass dynamically, baryon 
resonances occur abundantly, and quarks and gluons become confined in nucleons. This crossover 
process is controlled by the excited hadrons, as is schematically 
shown in the generic QCD phase diagram in Fig.~\ref{universe}. In this process 
strong QCD (sQCD) is born as the theory describing the interaction of colored quarks and gluons. These are the 
phenomena that we are exploring at Jefferson Lab and other accelerators around the world - the full 
discovery of the baryon (and meson) spectrum, the role of chiral symmetry breaking and the 
generation of dynamical quark mass in confinement. While we cannot recreate in the laboratory the 
exact condition that occurred during this period in the universe, with existing accelerators we can 
explore these processes in relative isolation. With electron machines and high energy photon beams 
in the few GeV energy range we search for undiscovered nucleon and baryon excitations.  

As the universe expands and cools down the coupling of quarks to the gluon field becomes stronger and 
quarks become more massive and form excited states in abundance.  This eventually leads  
to the forming of stable nucleons.

 \section{The quest for the missing baryon states}     
\label{missing-baryons}
The excited states of the nucleon have been studied experimentally 
since the 1950's~\cite{Anderson:1952nw}. They contributed to the discovery of the 
quark model in 1964 by 
Gell-Mann~\cite{GellMann:1964nj} and Zweig~\cite{Zweig:1981pd}, and were critical for the discovery 
of "color" degrees of freedom as introduced by Greenberg~\cite{Greenberg:1964pe}. 
The quark structure of baryons resulted in the prediction of a wealth of excited states  
with underlying spin-flavor and orbital symmetry of $SU(6) \otimes O(3)$, and led to 
a broad experimental effort to search for these states. Most of 
the initially observed states were found with hadronic probes. However, of the many excited 
states predicted in the quark model, only a fraction have been observed to date.  

It is interesting to point out recent findings that relate the observed baryon spectrum of different 
quark flavors with the baryon densities in the freeze out temperature in heavy ion collisions,
which show evidence for missing baryons in the strangeness and the charm baryon 
sector~\cite{Bazavov:2014xya,Bazavov:2014yba}. These data hint that an improved baryon model
including further unobserved light quark baryons would resolve the current discrepancy between
hot QCD lattice results and the results obtained using a baryon resonance model that includes only 
states listed by the PDG. A complete accounting of excited baryon states of all flavors seems 
essential for a quantitative description of the occurrence of baryons in the evolution of the 
microsecond old universe. It makes a systematic search for so far undiscovered nucleon states even 
more compelling.     

Search for the "missing" states and detailed studies of the resonance structure  
are now mostly carried out using electromagnetic probes and have been a major focus of  
hadron physics for the past two decades \cite{Burkert:2004sk}. A broad  
experimental effort has been underway with measurements 
of  exclusive meson photoproduction and electroproduction reactions, including 
many polarization observables. Precision data and the development of multi-channel 
partial wave analysis procedures have resulted in the discovery of several new excited states of the 
nucleon, which have been entered in the Review of Particle Physics (RPP)~\cite{Patrignani:2016xqp}, 
and additional ones may be entered in subsequent editions.  

 The importance and impact of nucleon spectroscopy for sQCD may be compared with 
 the impact that atomic spectroscopy had on the development of QED. It is through a complete 
 description of the entire atomic spectroscopy and small effects, such as the Lamb shift, that 
 QED is considered as fully established. In analogy, sQCD must be able to predict 
 the nucleon spectrum, its pole structure and exact energy dependences before we can 
 claim that the problem has been solved and we understand the spectrum. Of course, this in 
 turn requires that the nucleon spectrum is experimentally fully established, a charge to 
 us experimentalists to do our part. It requires a "global" approach, employing different
  experimental equipments and beams, and a systematic search for undiscovered 
  baryon states.        
  A quantitative description of baryon spectroscopy and the structure of excited 
nucleons must eventually involve solving QCD for  
a complex strongly interacting multi-particle system. 
Recent advances in Lattice QCD led to predictions of the nucleon spectrum in QCD with 
dynamical quarks~\cite{Dudek:2012ag}, albeit with still large pion 
 masses of 396 MeV. At the present time predictions can therefore only be taken as indicative of the 
 quantum numbers of excited states and not of the energy levels and pole structure of specific states. In parallel, 
 the development of dynamical coupled channel models is being pursued with new vigor. 
 The EBAC group at JLab has confirmed~\cite{Suzuki:2009nj} that dynamical effects can result 
 in significant mass shifts of the excited states. As a particularly striking result, a very large 
 shift was found for the Roper resonance pole mass to $\approx$ 1360 MeV downward from its bare 
 core mass of 1736 MeV. This result has clarified the longstanding puzzle of the incorrect 
 mass ordering of $N(1440){1/2}^+$ and $N(1535){1/2}^-$ resonances in the constituent 
 quark model. Developments on the phenomenological side go hand in hand with a 
 world-wide experimental effort to produce high precision data in many different channel 
 as a basis for a determination of the light-quark baryon 
 resonance spectrum. 
 On the example of experimental results from CLAS, the strong impact of precise meson 
 photoproduction data is discussed.  
 Several reviews have recently been published on the baryon 
 spectrum and structure of excited states~\cite{Klempt:2009pi,Tiator:2011pw,Aznauryan:2011qj,Aznauryan:2012ba,Crede:2013sze,Mokeev:2015lda}, 
 and on the 50 years puzzle of the Roper resonance~\cite{Burkert:2017djo}. 

Accounting for the complete excitation spectrum of the nucleon (protons and neutrons) 
and understanding the effective degrees of freedom is among the most important and  
certainly the most challenging task of hadron physics. The experimental N* program currently 
focusses on the search for new excited states in the light-quark sector of $N^*$ and $\Delta^*$ states and in 
the mass range up to 2.5 GeV  using energy-tagged photon beams in the few GeV range. Employing meson 
electroproduction  the study of the internal
structure of prominent resonances has been another major focus of the experimental exploration with CLAS.  
%\begin{figure}[h]
%\centering
%\resizebox{1.0\columnwidth}{!}{\includegraphics{CLAS-channels.png}}
%\caption{Experimental program with CLAS. The table shows observables that may be measured in 
%photoproduction of mesons from proton and neutrons. The columns indicate the observables of unpolarized, 
%beam polarized, target polarized and recoil polarization for single and double polarization experiments. The checkmarks indicate the status of the analysis and the publication.}
%\label{CLAS-channels}
%\end{figure}
\section{Completing the $N^*$ and $\Delta^*$ Spectrum}
\label{sec:1}
The complex structure of the light-quark baryon spectrum complicates the experimental 
search for individual states. As a consequence of the strong interaction, resonances are wide, 
often 200 MeV to 350 MeV, 
and are difficult to be uniquely identified when only differential cross sections are measured. Most of the excited 
nucleon states listed in the Review of Particle Physics prior to 2010 have been observed in elastic pion scattering 
$\pi N \to \pi N$. However there are important limitations in the sensitivity to the 
higher mass nucleon states that may have small $\Gamma_{\pi N}$ decay widths.  
The extraction of resonance contributions then becomes exceedingly difficult in $\pi N $ scattering.

Estimates for alternative decay channels have 
been made in quark model calculations~\cite{Capstick:1993kb} for various channels.  This has
 led to a major experimental effort at JLab, ELSA, GRAAL, and MAMI, LEPS and other laboratories to chart differential cross sections and polarization observables for a variety of meson
 photoproduction channels. At JLab with CLAS, many final states have been measured with high 
 precision~\cite{Dugger:2005my,Dugger:2009pn,Williams:2009yj,Williams:2009aa,Williams:2009ab,Bradford:2006ba,Bradford:2005pt,McCracken:2009ra,Dey:2010hh,McNabb:2003nf,Paterson:2016vmc,Compton:2017xkt,Ho:2017kca} and are now employed in multi-channel analyses. 
\begin{figure}[t]
\centering
\resizebox{0.85\columnwidth}{!}{\includegraphics{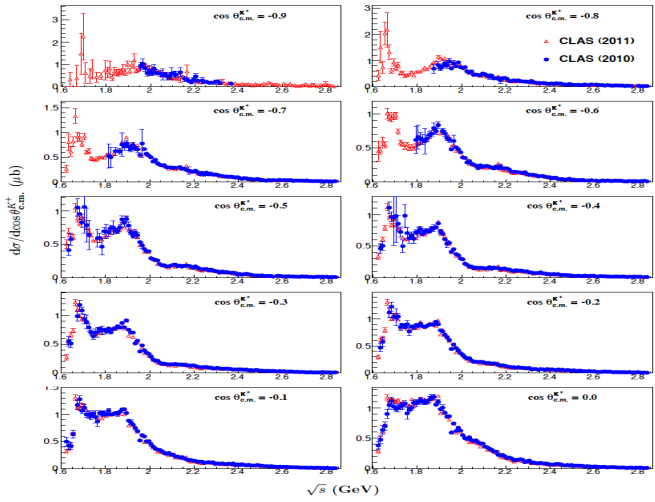}}
\caption{Invariant K$\Lambda$ mass dependence of differential cross sections in bins of $\rm\cos{\theta^K_{c.m.}}$ }
\label{KLambda-crs}
\end{figure}

\begin{figure}[h]
\centering
\resizebox{0.80\columnwidth}{!}{\includegraphics{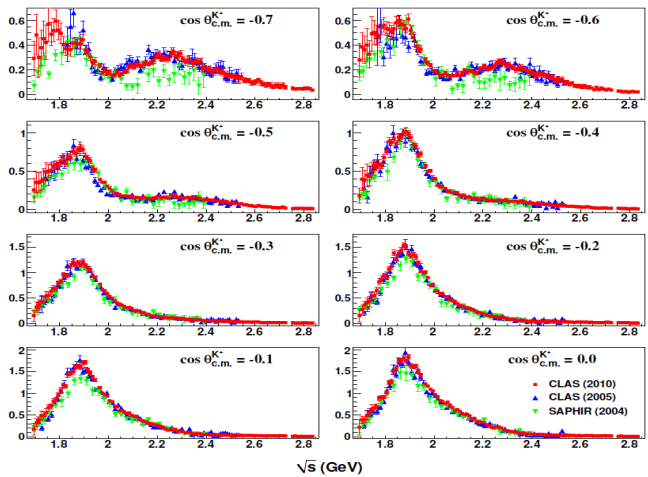}}
\caption{Invariant mass dependence of the $\gamma p \to K^+\Sigma^\circ$ differential cross section in the backward
polar angle range. }
\label{KSigma-crs}
\end{figure}
\subsection{New excited nucleon states from open strangeness photoproduction}
 \label{KLambda}
 In the past decade one focus has been on measurements of $\gamma p \to K^+\Lambda$, using a 
 polarized photon beam several polarization observables can be measured by analyzing the 
 parity violating decay of the recoil $\Lambda \to p \pi^-$. It is well known that the energy-dependence of 
 a partial-wave amplitude for one particular channel is influenced by other reaction 
channels due to unitarity constraints. To fully describe the energy-dependence 
of an amplitude one has to include other reaction channels in a coupled-channel approach. 
Such analyses have been developed by the Bonn-Gatchina group~\cite{Anisovich:2011fc}, 
at JLab~\cite{JuliaDiaz:2007kz}, Bonn-J\"ulich~\cite{Ronchen:2014cna}, Argonne-Osaka~\cite{Kamano:2013iva}, and other groups.

The data sets with the highest impact on resonance amplitudes in the mass range above 1.7~GeV have been 
kaon-hyperon production using a spin-polarized photon beam and 
where the polarization of the $\Lambda$ or $\Sigma^\circ$ is also measured. The high precision cross section and 
polarization data~\cite{Bradford:2006ba,Bradford:2005pt,McCracken:2009ra,Dey:2010hh,McNabb:2003nf} provide 
nearly full polar angle coverage and span the $K^+\Lambda$ invariant mass range 
from threshold to 2.9 GeV, hence covering the full nucleon resonance domain where new states might be discovered. 

The backward angles $K^+\Lambda$ data in Fig.\ref{KLambda-crs} show clear resonance-like structures 
at 1.7 GeV and 1.9 GeV that are particularly prominent and well-separated from other structures, while 
at more forward angles (not shown) t-channel processes become prominent and dominate the cross section.
The broad enhancement at 2.2~GeV may also indicate resonant behavior although it is less visible at more 
central angles with larger background contributions. 
The $K^+\Sigma$ channel also indicates significant resonant behavior as seen in Fig.~\ref{KSigma-crs}. The peak structure at 1.9 GeV is 
present at all angles with a maximum strength near 90 degrees, consistent with the behavior of a $J^P= {3/2}^+$ 
p-wave. Other structures near 2.2 to 2.3~GeV are also visible. 
Still, only a full partial wave analysis can determine the underlying resonances, their masses and spin-parity.  
The task is somewhat easier for the $K\Lambda$ channel, as the iso-scalar nature of the $\Lambda$ selects 
isospin-${1\over 2}$ states to contribute to the $K\Lambda$ 
final state, while both isospin-${1\over 2}$ and isospin-${3\over 2}$ states can contribute to the $K\Sigma$ final state.     

\begin{figure}[b]
\centering
\resizebox{0.85\columnwidth}{!}{\includegraphics{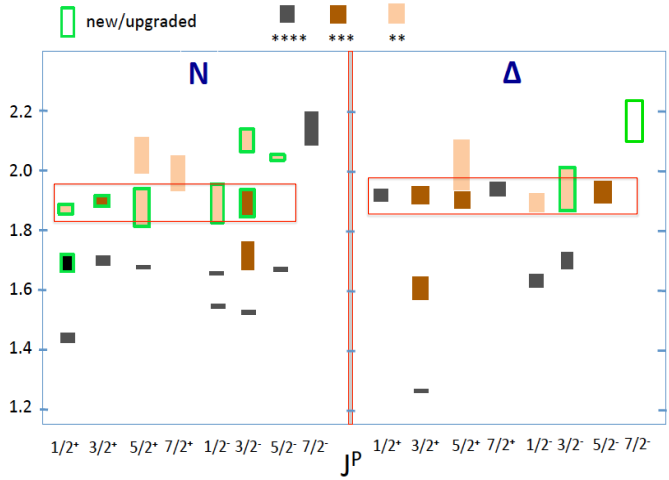}}
\caption{Nucleon and $\Delta$ resonance spectrum up to 2.2 GeV in RPP 2016~\cite{Patrignani:2016xqp}. The
new states and states with improved evidence observed in the recent Bonn-Gatchina multi-channel analysis are 
shown with the green frame. The red frames highlight the apparent mass degeneracy of five or six states with 
different spin and parity.  The analysis includes all the $K^+\Lambda$ and $K^+\Sigma^\circ$ cross section and 
polarization data.}
\label{pdg2016}
\end{figure}
%\begin{figure}[h]
%\centering
%\resizebox{0.60\columnwidth}{!}{\includegraphics{omega_fit.pdf}}
%\caption{Phase motion of the partial wave fit to the $\gamma p \to p \omega$ differential cross section and spin density matrix elements. 3 resonant 
%states, the subthreshold resonance $N(1680){5\over 2}^+$, $N(2190){7\over 2}^-$, and the missing $N(2000){5\over 2}^+$ are needed to fit the data (solid line). Fits without $N(2000){5\over 2}^+$ (dashed-dotted line), or without $N(1680){5\over 2}^+$ (dashed line) cannot reproduce the data.}
%\label{omega}
%\end{figure}

These cross section data together with the $\Lambda$ and $\Sigma$ recoil polarization and polarization transfer data 
to the $\Lambda$ and $\Sigma$ had strong impact on the discovery of several new nucleon 
states. They also provided new evidence for several candidate states that had been observed previously but lacked 
confirmation, as shown in Fig.~\ref{pdg2016}. It is interesting to observe that five of the observed nucleon states have nearly 
degenerate masses near 1.9~GeV. 
Similarly, the new $\Delta$ state appears to complete a mass degenerate multiplet near 1.9~GeV as well. There is no obvious 
mechanism for this apparent degeneracy.  Nonetheless, all new states may be accommodated within the symmetric 
constituent quark model based on $SU(6)\otimes O(3)$ symmetry group as far as quantum numbers are concerned. As discussed 
in section~\ref{intro} for the case of the Roper resonance $N(1440){1\over 2}^+$, the masses of all pure  quark model states 
need to be corrected for dynamical coupled channel effects to compare them with observed resonances.  The same applies to 
the recent Lattice QCD predictions~\cite{Edwards:2011jj} for the $N^*$ and $\Delta^*$ spectra. 

\subsection{ A new high-mass isospin 3/2 state confirmed in the $N\pi$ final state}  
\begin{figure}[h]
\centering
\resizebox{0.7\columnwidth}{!}{\includegraphics{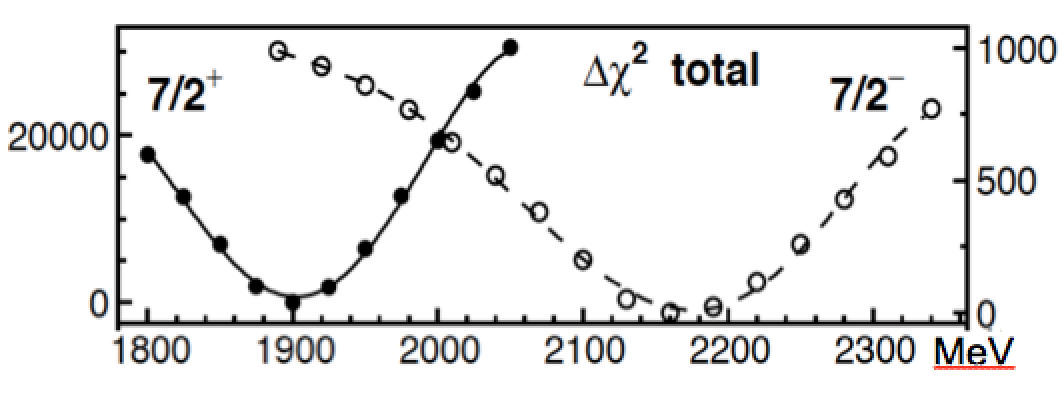}}
\caption{Evidence for $\Delta(2200){7/2}^-$ from $\gamma p \to N\pi$ differential cross sections and single and double polarization 
measurements  at CLAS and CBELSA. Evidence has also been observed in $K\Sigma$ and $p\pi^\circ\eta$ final states.}
\label{Delta2200}
\end{figure}

The power of polarization measurements has been demonstrated 
in the strong evidence seen for the $\Delta(2200){7/2}^-$ state. Although the state couples to $N\pi$ with a branching 
ratio of just 3.5\%, the combination of  precise differential cross section and single and double polarization 
measurements made this possible~\cite{Anisovich:2015gia}. Figure~\ref{Delta2200} shows 
the mass scan for the well known $\Delta(1950)7/2^+$ and the new $\Delta(2200){7/2}^-$ showing clear effects on 
the effective $\Delta\chi^2$. The state had prior only a one-star rating. Its empirical mass value is indicated in 
Fig.~\ref{pdg2016} with the open green frame. 
\subsection{Vector meson photoproduction}
\label{Vectormeson}
\begin{figure}[b]
\centering
\resizebox{0.7\columnwidth}{!}{\includegraphics{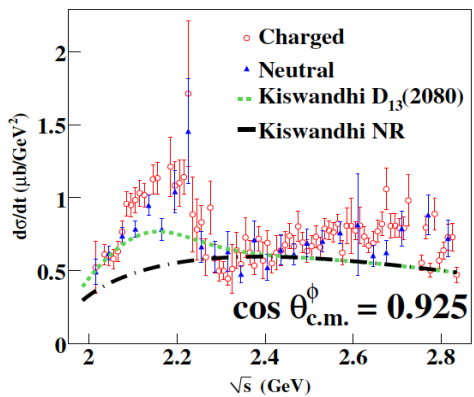}}
\caption{Differential cross sections of $\gamma p \to p \phi$  production for the most forward angle bin. The two curves 
refer to fits without (dashed) and with (dotted) a known resonance at 2.08 GeV included. }
\label{phi_forward}
\end{figure}
%\vspace{-1cm}
In the mass range above 2.0~GeV resonances tend to decouple from simple 2-body final states like $N\pi$, $N\eta$, and 
$K\Lambda$. We have to consider more complex final states with multi-mesons, such as $N\pi\pi$ and $N\pi\eta$, as well as 
vector mesons $N\omega$, $N\phi$, and $K^*\Sigma$. The study of such final states adds significant 
complexity as more amplitudes can contribute to photoproduction of spin-1 mesons, compared to pseudo-scalar meson 
production. As is the case for $N\eta$ production, 
the $N\omega$ channel is selective to isospin $1\over 2$ nucleon states.       
CLAS has collected a tremendous amount of data in the $p\omega$~\cite{Williams:2009aa,Williams:2009ab,Collins:2017vev}, 
$p\phi$~\cite{Seraydaryan:2013ija,Dey:2014tfa}, and $K^*\Sigma$~\cite{Anisovich:2017rpe} final states on differential cross sections and spin-density matrix elements, that are now entering 
into the more complex multi-channel analyses such as Bonn-Gatchina. 
The CLAS collaboration performed a single channel event-based analysis, and provide further evidence for the $N(2000){5/2}^+$.  

Photoproduction of $\phi$ mesons is also considered a potentially rich source of new excited nucleon states in the mass range above 2 GeV. 
Some lower mass states such as $N(1535){1/2}^-$ below the $N\phi$ threshold may have significant $s\bar{s}$ components~\cite{Liu:2005pm}. Such components 
may result in states coupling to $p\phi$ with significant strength above threshold. Differential cross sections and spin-density matrix elements 
have been measured for $\gamma p \to p \phi$ in a mass range up to nearly 3 GeV. 
%In Fig.~\ref{phi} structures are seen near 2.2~GeV in the forward most angle bins and at very backward angles for both decay channels $\phi \to K^+K^-$ and $\phi \to K_l^0K_s^0$, and with the exception of the smallest forward angle bin the structures are more prominent at backward angles. 
A multi-channel partial wave analysis is required to pull out any significant resonance strength in this channel. 
Figure~\ref{phi_forward} shows the differential cross section $d\sigma /dt$ 
of the most forward angle bin. A broad structure at 2.2 GeV is present, but does not show the typical Breit-Wigner behavior of a single 
resonance. It also does 
not fit the data in a larger angle range, which indicates that contributions other than genuine resonances may be significant. 
The forward and backward angle structures 
may also hint at the presence of dynamical effects possibly due to molecular contributions such as diquark-anti-triquark 
contributions~\cite{Lebed:2015dca}, the strangeness equivalent to the recently observed hidden charm $P_c^+$ states.   

Another process that has promise in the search for new excited baryon states, including those with 
isospin-${3/2}$, is $\gamma p \to K^*\Sigma$. 
In distinction to the vector mesons discussed above, diffractive processes do not play a role in this channel, 
which then may allow more direct access to s-channel resonance production.    

\begin{figure}[t]
\centering
\resizebox{0.75\columnwidth}{!}{\includegraphics{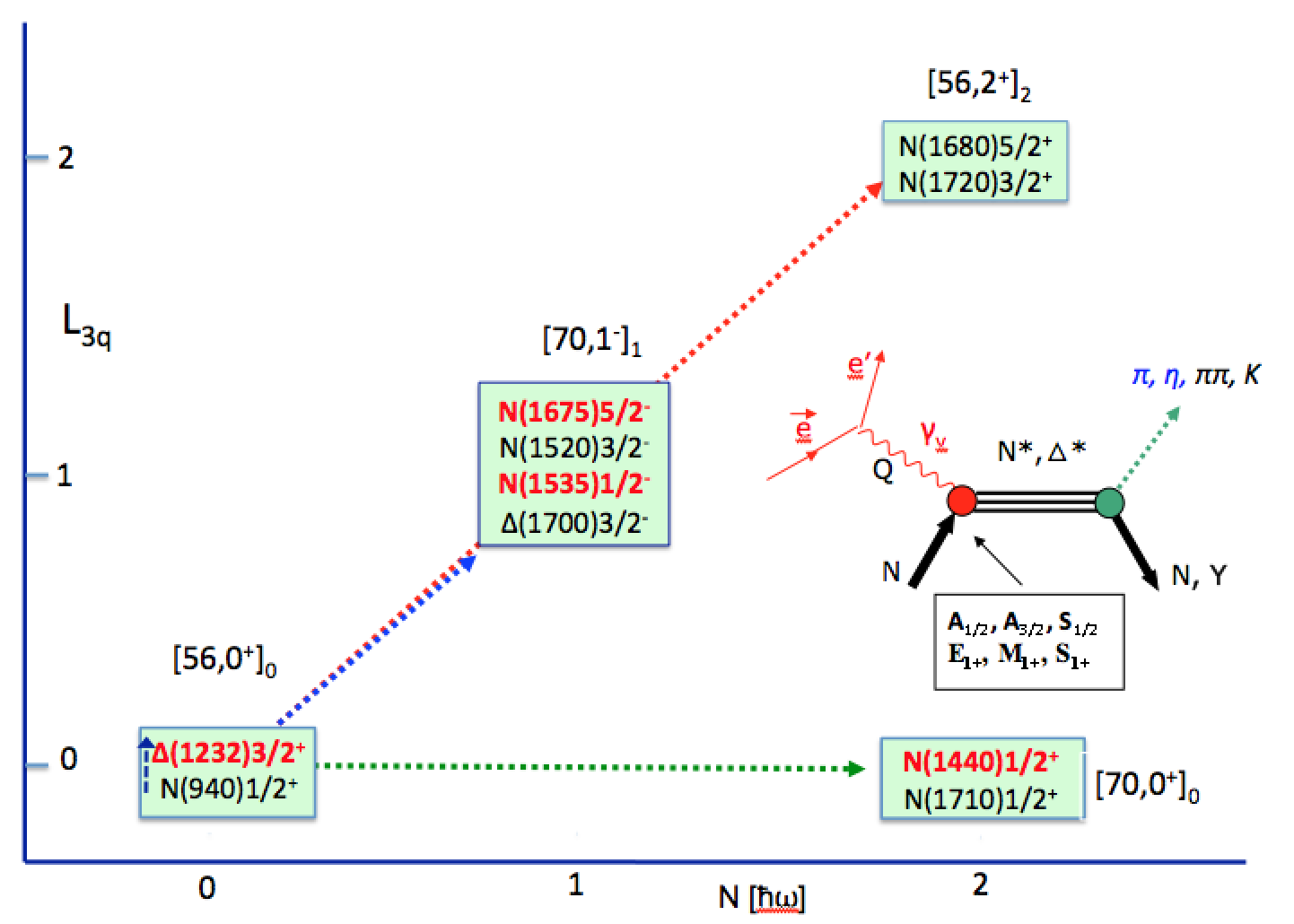}}
\caption{Schematic of $SU(6)\otimes O(3)$ supermultiplets with selected prominent excited states that have been explored in $e p \to e^\prime \pi^+ n$, 
$e p \to e^\prime p^\prime \pi^\circ$ and $ep \to e^\prime p^\prime \pi^+\pi^-$. Only the states highlighted in red are discussed here.
The insert shows the helicity amplitudes and electromagnetic multipoles  extracted from the data. }
\label{su6}
\end{figure}

We can conclude that meson photoproduction has become an essential tool in the search for new excited baryons. 
The exploration of the internal structure 
of excited states and the effective degrees of freedom contributing to s-channel resonance excitation requires the use of electron beams,  
where the virtuality $Q^2$ of the exchanged photon can be varied to probe the spatial structure (Fig.~\ref{su6}). This is discussed in
the following section. 

\section{Structure of excited nucleons}
\label{structure}

\begin{figure}[t]
\centering
\resizebox{0.95\columnwidth}{!}{\includegraphics{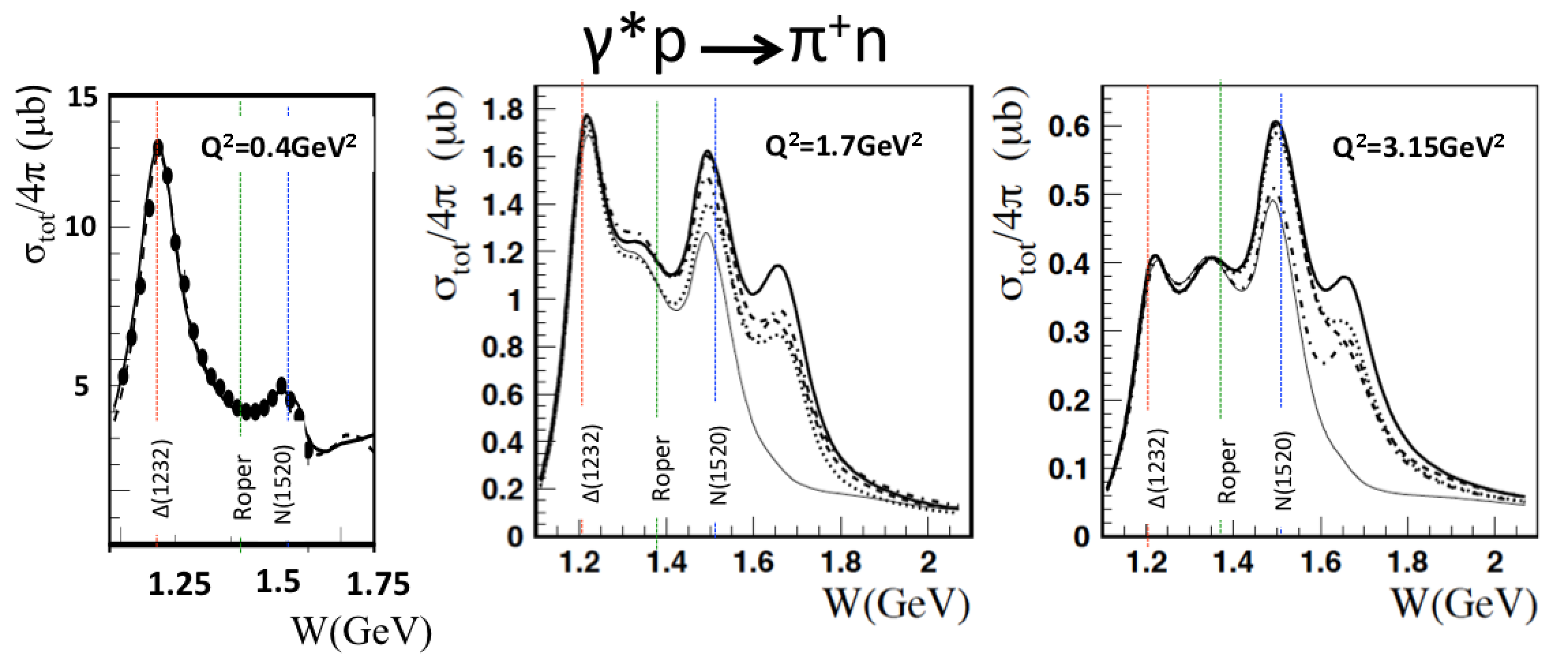}}
\caption{ Evolution of the resonance strength with $Q^2$. The 3 panels show the strength of the 4 enhancements, which
are related dominantly to certain resonances, vary with increasing $Q^2$ significantly.  }
\label{resonance_evolution}
\end{figure}

This will enable us to draw some conclusions about the effective degrees of freedom underlying the resonance transition strength. 
The fact that resonance can exhibit very different $Q^2$-dependencies in their respective helicity amplitudes is demonstrated with the 3 panels in
in~Figure~\ref{resonance_evolution}  
where integrated cross sections are displayed taken at different photon virtuality $Q^2$. They exhibit a number 
of enhancements that are associated with several prominent resonance, the $\Delta(1232){3/2}^+$, the Roper 
$N(1440){1/2}^+$, $N(1520){3/2}^-$, and $N(1680){5/2}^+$. The strength of the $\Delta$ excitation seen at small 
$Q^2$ drops rapidly at higher $Q^2$. The Roper $N(1440){1/2}^+$ is not visible at low 
$Q^2$ but emerges as $Q^2$ increases. The $N(1520){3/2}^-$ and $N(1535){1/2}^-$ bump is small at low $Q^2$ and 
becomes the dominant peak at highest $Q^2$. This     
 $Q^2$ dependence shows that the various underlying resonances behave differently with the increase in $Q^2$, which 
 is indicative of different degrees-of-freedom underlying their respective excitation strengths. 
 
\begin{figure}[t]
\centering
\resizebox{0.95\columnwidth}{!}{\includegraphics{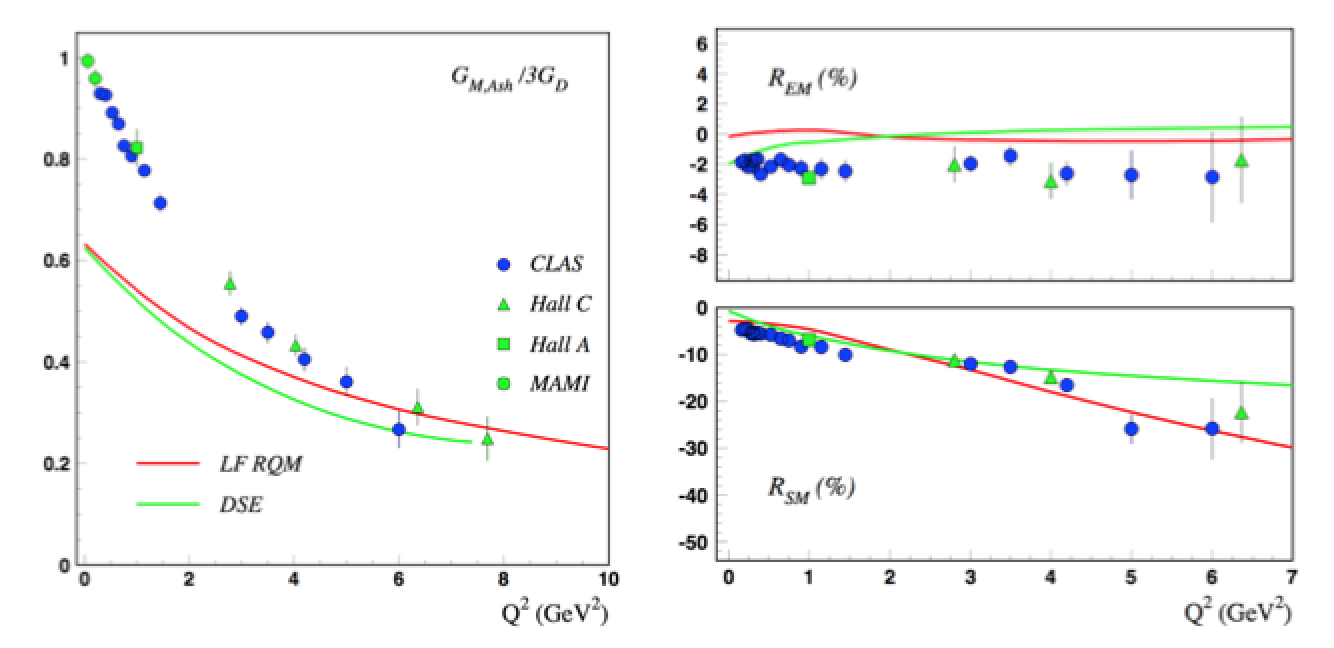}}
\caption{ Left panel: $N\Delta$ transition magnetic form factor. Right panel: Electric quadrupole ratio $R_{EM}$ (top), and 
scalar quadrupole ratio $R_{SM}$ (bottom). }
\label{Delta}
\end{figure}

Electroproduction of final states with pseudoscalar mesons 
(e.g. $N\pi$, $p\eta$, $K\Lambda$) have been employed with CLAS, leading to new insights into the dependence of 
effective degrees of freedom on the distance scale, e.g. meson-baryon, constituent quarks, dressed quarks, and bare quark 
contributions. Several excited states, shown 
in Fig.~\ref{su6} assigned to their primary $SU(6) \otimes O(3)$ supermultiplets, have been studied. The $p\Delta^+(1232){3/2}^+$
 transition is now well measured in a large range of $Q^2$~\cite{Joo:2001tw,Ungaro:2006df,Frolov:1998pw,Aznauryan:2009mx}. 
The transition amplitudes, characterizing the $N-\Delta(1232)$ transition, are usually defined as the magnetic transition form 
 factor $G_M^\Delta$, the electric quadrupole ratio $R_{EM} = E_{1+}/M_{1+}$  and the scalar quadrupole ratio $R_{SM} = S_{1+}/M_{1+}$.
The current status of these quantities are shown in Fig.~\ref{Delta}. 
The data are compared to two recent calculations, one based on the LF/RQM~\cite{Aznauryan:2016wwm}, and on the 
DSE/QCD approach~\cite{Segovia:2014aza}. For the magnetic transition form factor both calculations are close to each 
other, and agree with the data  at the high $Q^2$ end.  Both calculations project very small $R_{EM}$ quark contributions 
throughout the measured $Q^2$ range.  They show similar trends for $R_{SM}$ at low and medium $Q^2$, but are diverging 
at the highest $Q^2$. Extending the data to even higher $Q^2$ should be revealing. Asymptotic QCD predicts a 
constant value for  $R_{SM}$, while holographic QCD models predict a specific limit of $R_{SM}(Q^2 \to \infty) \to -1$~\cite{Grigoryan:2009pp}.   

\begin{figure}[h]
\centering
\resizebox{0.45\columnwidth}{!}{\includegraphics{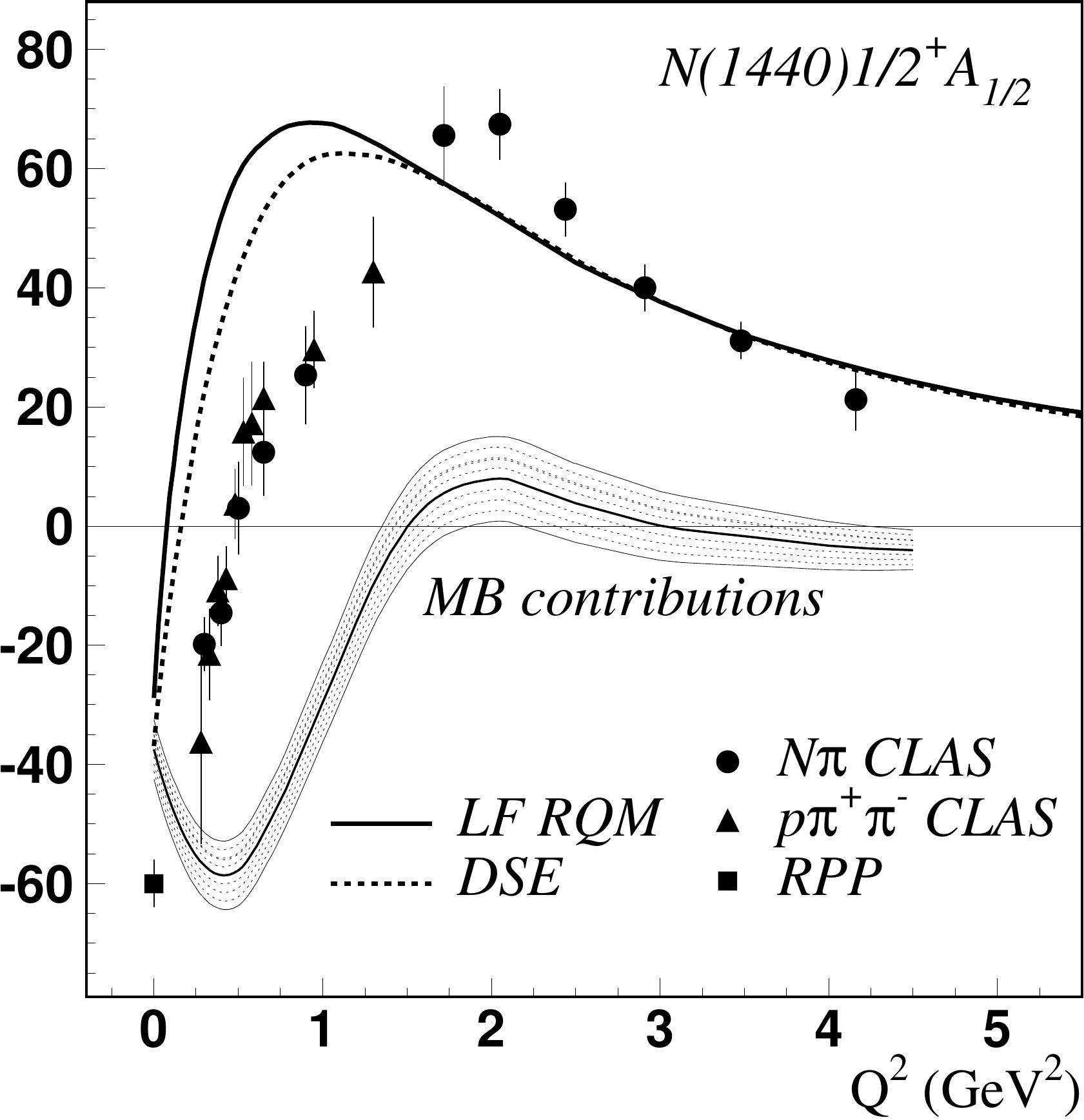}}
\hspace{0.45cm}\resizebox{0.46\columnwidth}{!}{\includegraphics{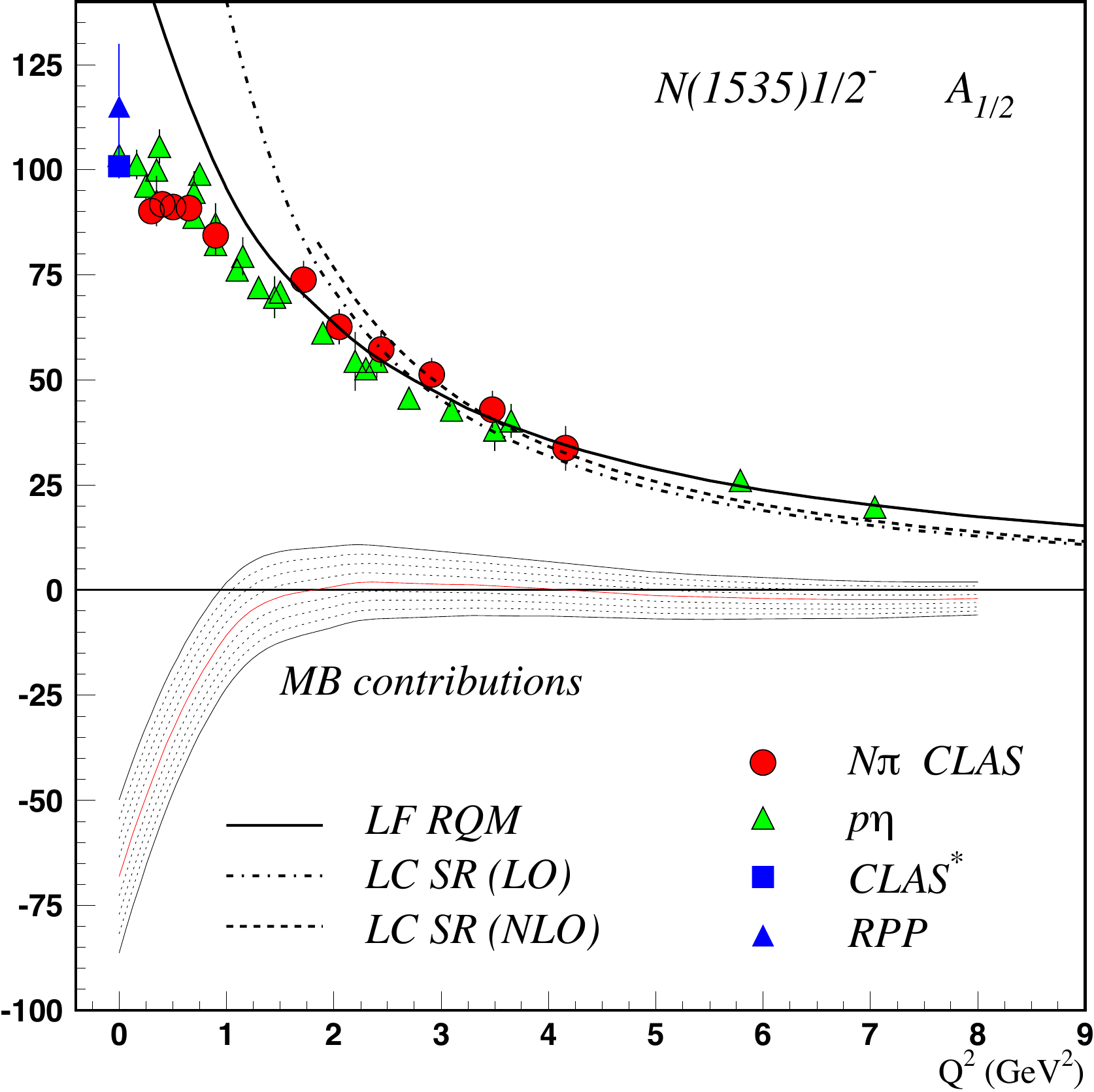}}
\caption{Left panel: The transverse helicity amplitudes $A_{1/2}$ for the Roper resonance $N(1440){1/2}^+$. 
Data are from CLAS compared
to the LF RQM with momentum-dependent quark masses and with projections from the DSE approach. The 
dashed band indicates size of non 3-quark contributions obtained from a the difference of the LF RQM curve 
and the CLAS data. The right panel shows the $A_{1/2}$ amplitude for the $N(1535){1/2}^-$ compared to  
LF RQM calculations and to lattice QCD-based Light Cone Sum Rule calculation in LO and NLO approximation~\cite{Anikin:2015ita}.}
\label{p11_s11}
\end{figure}

Two of the prominent higher mass states, the Roper resonance
$N(1440){1/2}^+$  and $N(1535){1/2}^-$ are shown in Fig.~\ref{p11_s11} as representative 
examples~\cite{Aznauryan:2009mx,Aznauryan:2008pe} from a wide program at 
JLab~\cite{Mokeev:2012vsa,Denizli:2007tq,Armstrong:1998wg,Egiyan:2006ks,Park:2007tn,Park:2014yea}.  For these two states advanced
relativistic quark model calculations~\cite{Aznauryan:2015zta} and QCD-linked calculations from 
 \begin{figure}[h]
\centering
\resizebox{0.75\columnwidth}{!}{\includegraphics{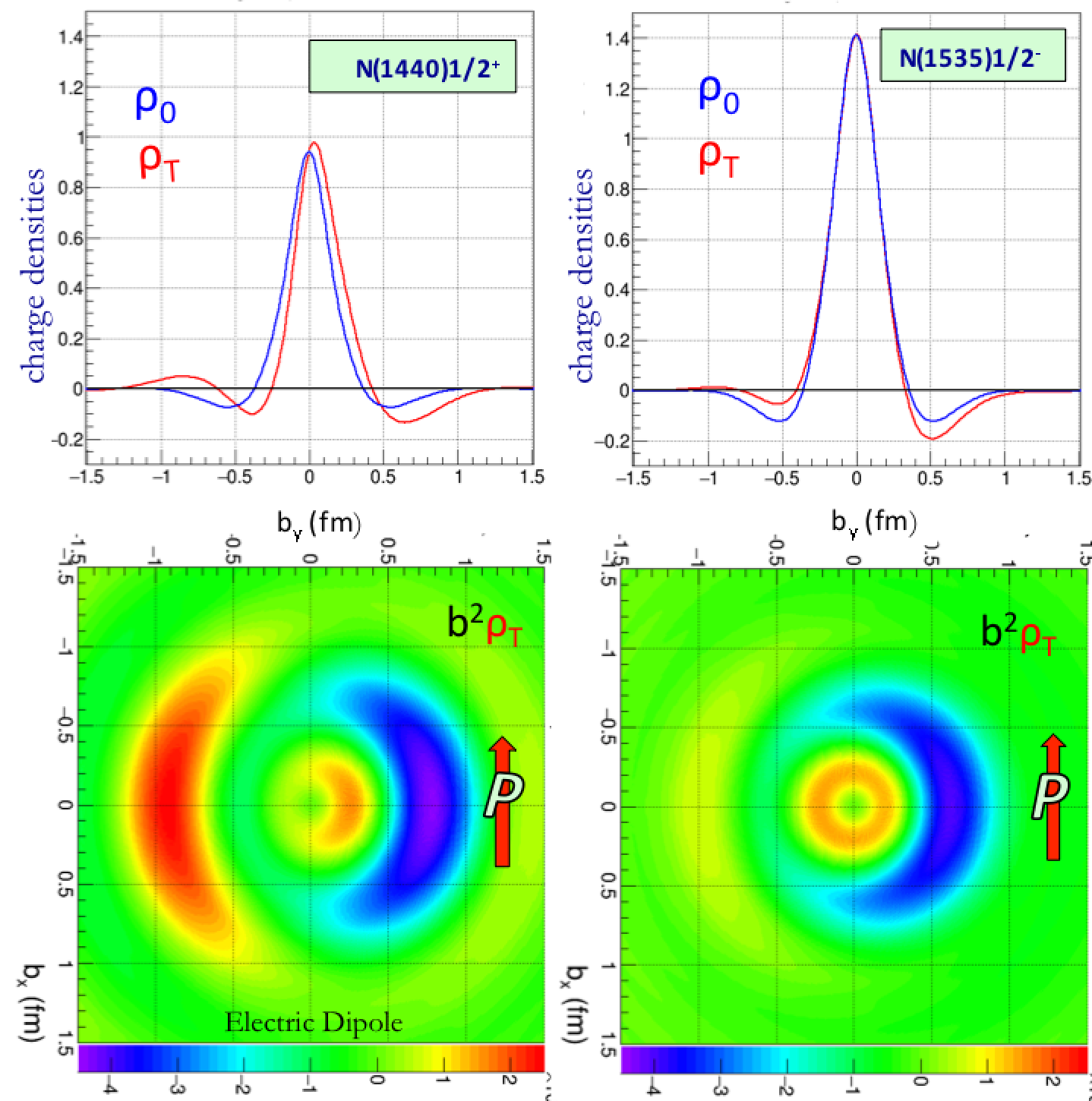}}
\caption{Charge densities for the two resonances. 
Left panels: $N(1440){1/2}^+$, top: projection of transition charge densities on $b_y$, bottom: transition charge densities when the proton  
is spin polarized along $b_x$. Right panels: same for $N(1535){1/2}^-$. Note that the densities are scaled with 
$b^2$ to emphasize the outer wings. Color code:negative charge tends to blue, positive charge tends to red. For ease of 
comparison all scales are the same. Figures courtesy of F.X. Girod. }
\label{charge_densities}
\end{figure}
Dyson-Schwinger equations~\cite{Segovia:2015hra} as well as Light Cone sum rule (LCSR)~\cite{Anikin:2015ita} have become available, 
for the first time employing QCD-based modeling of the excitation 
of the quark core. There is near quantitative agreement of both calculations with the data at $Q^2 > 1.5$~GeV$^2$. Note that
the LF RQM includes a momentum-dependent quark mass parameterization that is fixed to describe the nucleon electromagnetic 
form factors. The same function is used for all transition amplitudes. 
This result strongly indicates that at the scale of the quark core the Roper resonance is the first radial excitation of the 
nucleon. From the excellent agreement with LF RQM and the LC SR approaches we can also draw the conclusions the 
the $N(1535){1/2}^-$ resonance as its core is the first orbital excitation of the nucleon. .  
We want to emphasize, however, it is only from the measurement of the excitation strength at high enough $Q^2$ that we 
can draw such conclusions~\cite{Burkert:2017djo,Aznauryan:2011qj}, while the peripheral behavior at low $Q^2$ requires 
the inclusion of hadronic degrees-of-freedom for a quantitative description.   
For the Roper resonance such contributions have been described successfully in dynamical meson-baryon 
models~\cite{Obukhovsky:2011sc} and in effective field theory~\cite{Bauer:2014cqa}. 

Knowledge of the helicity amplitudes in a large $Q^2$ range allows for the determination of the transition charge densities on the light 
cone in transverse impact parameter space ($b_x, b_y$)~\cite{Tiator:2008kd}. Figure~\ref{charge_densities} shows the comparison 
of $N(1440){1/2}^+$ and  $N(1535){1/2}^-$. There are clear differences in the charge transition densities between 
the two states. The Roper resonance has a softer positive core and a wider negative outer cloud than $N(1535)$.  
It also exhibits a larger shift in $b_y$ when the transition is from a proton that is polarized along the $b_x$ axis. Both transitions 
show an electric transition dipole moment, the one of the Roper appears as significantly stronger and shows a more pronounced 
charge asymmetry. 

As these transition charge densities represent moments of transition amplitudes they may be accessible to LQCD 
and other implementations of sQCD.     
 \begin{figure}[t]
\centering
\vspace{1cm}
\resizebox{0.95\columnwidth}{!}{\includegraphics{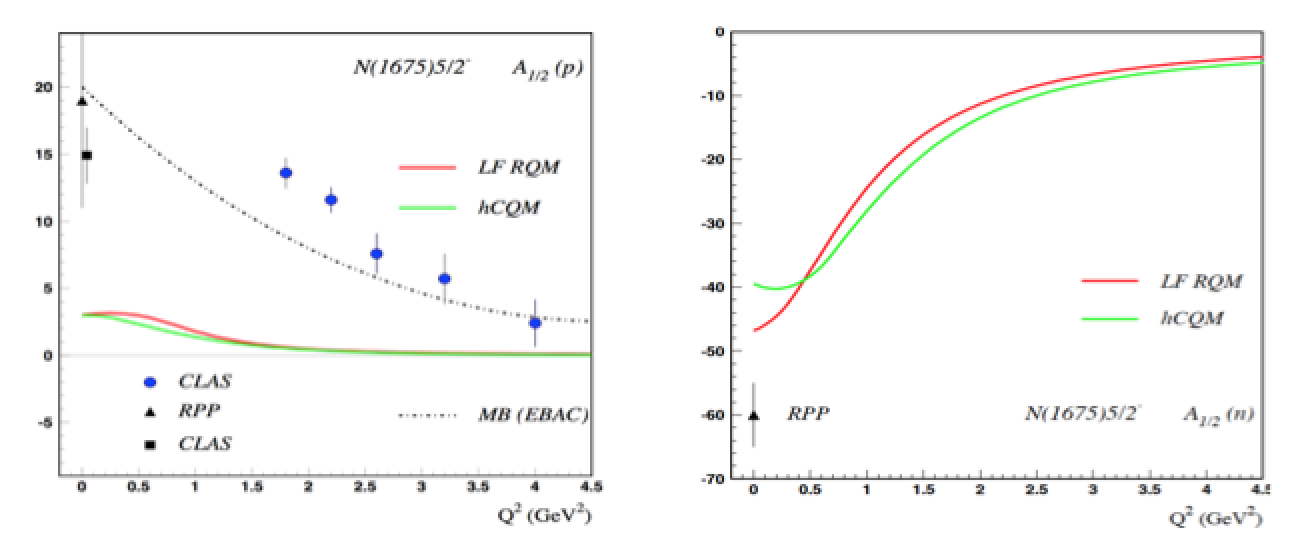}}
\caption{Helicity amplitude $A_{1/2}$ for $N^+(1675){5/2}^-$ off proton target (left), and for $N^0(1675){5/2}^-$ off neutron 
target (right). 
 }
\label{N1675}
\end{figure}

\subsection{The $N(1675){5/2}^-$  state - revealing the meson-baryon contributions}
In previous discussions we have assumed that meson-baryon degrees of freedom provide significant strength to the 
resonance excitation in the low $Q^2$ domain where quark the based approaches LF RQM, DSE/QCD, and LCSR calculations fail to 
reproduce the transition amplitudes quantitatively. Our conclusion rests, in part, with this assumption. 
But, how can we be certain of the validity of this assumption? 

The $N(1675){5/2}^-$ resonance allows testing this assumption, quantitatively. Figure~\ref{N1675} shows 
our current knowledge of the transverse helicity amplitude $A_{1/2}(Q^2)$ for the proton and the neutron 
and LF RQM~\cite{Aznauryan:2017nkz} and 
hypercentral CQM~\cite{Santopinto:2012nq} calculations. The specific quark transition for a $J^P = 5/2^-$ state belonging to the 
$SU(6)\otimes O(3)] = [70, 1^-]$ supermultiplet configuration prohibits the transition from the proton in a single quark transition.
This suppression is known as the Moorhouse selection rule~\cite{Moorhouse:1966jn}, and is valid for the transverse transition 
amplitudes $A_{1/2}$ and $A_{3/2}$ at all $Q^2$. 
It should be noted that this selection rule does apply only to the 
transition from protons but not from neutrons. Modern quark models, that go beyond single quark transitions,  
confirm quantitatively the suppression resulting in very 
small transition amplitudes from protons but large ones from neutrons. The measured helicity amplitudes off the protons are almost  
exclusively due to meson-baryon contributions as the dynamical coupled channel (DCC) calculation indicates (dashed line). 
The quark model prediction on the neutron predict large amplitudes at the photon point consistent with the single data point.
Note that the differences data-model for the proton and for the neutron have opposite signs but are of 
about the same magnitude of $\Delta A^p_{1/2}(0) =16\pm 8 \times 10^{-3}$GeV$^{-1}$, and 
$\Delta A^n_{1/2}(0) = -13\pm 5 \times 10^{-3}$GeV$^{-1}$.
A very similar behavior is seen for the $A_{3/2}$ amplitudes $\Delta A^p_{3/2}(0) =15\pm 5 \times 10^{-3}$GeV$^{-1}$, and 
$\Delta A^n_{3/2}(0) = -23\pm 10 \times 10^{-3}$GeV$^{-1}$~\cite{Aznauryan:2017nkz}. The close correlation of the 
DCC calculation and the measured data for the case when quark contributions are nearly absent,    
supports the phenomenological description of the helicity amplitudes in terms of a 3-quark core that dominate at high $Q^2$ and 
meson-baryon contributions that can make important contributions at lower $Q^2$. 

\section{Conclusions and Outlook}
Over the past several years, eight light-quark baryon states in the mass range from 1.85 to 2.25 GeV have been either discovered, or 
evidence for their existence has been brought close to certainty. To a large degree this is the result of adding 
very precise photoproduction data in open strangeness channels to the data base that is included in multi-channel partial wave 
analyses. The measurement of polarization observables in these processes has been 
critical. In the mass range above 2 GeV more complex processes such as vector mesons or $\Delta\pi$ 
may have sensitivity to states with higher masses and require more complex analyses techniques. 
Precision data in such channels have been available for a few years, but they have not been fully incorporated in 
multi-channel partial wave analyses processes. 

There has been progress to predict the nucleon spectrum from first principles within QCD on the lattice. While pion 
masses of about 400 MeV are still too large for precise predictions of resonance masses and poles, the predicted 
quantum numbers coincide with $SU(6)$ symmetry and states predicted within constituent quark models. 

The light-quark baryon spectrum is likely also populated with hybrid excitations~\cite{Dudek:2012ag}, where the gluonic 
admixtures to the wave function are dominating the excitation. These states appear with the same quantum numbers as 
ordinary quark excitations, and can only be isolated from ordinary states due to the $Q^2$ dependence of their helicity 
amplitudes~\cite{Li:1991yba}, which is expected to be quite different from ordinary quark excitations. To search for 
these new hybrid states, new electroproduction data especially at low $Q^2$~\cite{E12-16-010} are needed, with different 
final states and at masses above 2 GeV.   

On the theoretical side, we have seen the first calculation of the resonance transition helicity amplitudes and transition 
form factors for the case of the 
$\Delta(1232)3/2^+$, the Roper $N(1440){1/2}^+$, and the $N(1535){1/2}^-$ within QCD-linked approaches. Here we see 
agreement with data is in the range of $Q^2 > 2-3$~GeV$^2$.  We also have seen that newly discovered nucleon resonances fit into 
the spectrum projected from LQCD with their quantum, albeit not (yet) with their mass assignments.   

Despite the very significant progress made in recent years to further establish the light-quark baryon spectrum and explore 
the internal structure of excited states, much remains to be done. A vast amount of precision data that have already been 
collected, must be included in the multi-channel analysis frameworks, and many polarization data sets are still to be analyzed. 
There are new data on 2-pion electroproduction~\cite{Isupov:2017lnd} that will extend the mass range for the extraction of 
transition helicity amplitudes for high mass resonances. Of particular interest is here the $N(1900){3/2}^+$, which only 
recently has become a well established excited nucleon state. There are also upcoming experiments to study resonance 
excitations at much higher $Q^2$ and with higher statistical precision 
at Jefferson Lab with CLAS12~\cite{E12-09-003} that may begin to reveal the transition to the bare quark core contributions at 
short distances.   

Let me finally conclude, that the community is still on track of fulfilling Nathan Isgur's vision of a 30 year program to 
solve the puzzle of the baryon spectrum. 

\begin{acknowledgements}
I want to thank Inna Aznauryan for providing the figures on the transition amplitudes, which have led to new insight into the degrees 
of freedom underlying resonance excitations. I also thank Ralf Gothe, Viktor Mokeev and Craig Roberts for numerous
 discussions on the subjects discussed in this presentation, and F.X. Girod for providing the color graphics in Fig.~\ref{charge_densities}. 
This work was supported by the U.S. Department of Energy, Office of Science,
Office of Nuclear Physics, under Contract No. DE-AC05-06OR23177. 
\end{acknowledgements}

\end{document}